\documentclass{sig-alternate-05-2015}
\usepackage{graphicx}
\usepackage{balance}  % for  \balance command ON LAST PAGE  (only there!)
\usepackage{booktabs}

\usepackage{multirow}
\usepackage{caption}
\usepackage{todonotes}

\usepackage{subcaption}
\usepackage{algorithm,algpseudocode}
\usepackage{bm}

\usepackage{etoolbox}
\makeatletter
\patchcmd{\maketitle}{\@copyrightspace}{}{}{}
\makeatother

\begin{document}

% Copyright
\setcopyright{acmcopyright}
%\setcopyright{acmlicensed}
%\setcopyright{rightsretained}
%\setcopyright{usgov}
%\setcopyright{usgovmixed}
%\setcopyright{cagov}
%\setcopyright{cagovmixed}

% DOI
\doi{10.475/123_4}

% ISBN
\isbn{123-4567-24-567/08/06}

%Conference
%\conferenceinfo{XXX '17}{XXX , XXXX, XXX, XXX XXXX.}

%\acmPrice{\$XX.00}

%
% --- Author Metadata here ---
%\conferenceinfo{XXX}{'17  XXXX, XXX XXXX}
%\CopyrightYear{2007} % Allows default copyright year (20XX) to be over-ridden - IF NEED BE.
%\crdata{0-12345-67-8/90/01}  % Allows default copyright data (0-89791-88-6/97/05) to be over-ridden - IF NEED BE.
% --- End of Author Metadata ---

\title{SSH (Sketch, Shingle, \& Hash) for Indexing Massive-Scale Time Series}
%\subtitle{[Extended Abstract]
%\titlenote{A full version of this paper is available as
%\textit{Author's Guide to Preparing ACM SIG Proceedings Using
%\LaTeX$2_\epsilon$\ and BibTeX} at

\numberofauthors{2} %  in this sample file, there are a *total*
\author{
% 1st. author
\alignauthor
Chen Luo\\
%\titlenote{Dr.~Trovato insisted his name be first.}\\
       \affaddr{Department of Computer Science}\\
       \affaddr{Rice University}\\
       \affaddr{Houston, Texas}\\
       \email{cl67@rice.edu}
% 2nd. author
\alignauthor
Anshumali Shrivastava\\
%\titlenote{The secretary disavows
%any knowledge of this author's actions.}\\
       \affaddr{Department of Computer Science}\\
       \affaddr{Rice University}\\
       \affaddr{Houston, Texas}\\
       \email{anshumali@rice.edu}
% 3rd. author
%\alignauthor Lars Th{\o}rv{\"a}ld\titlenote{This author is the
%one who did all the really hard work.}\\
%       \affaddr{The Th{\o}rv{\"a}ld Group}\\
%       \affaddr{1 Th{\o}rv{\"a}ld Circle}\\
%       \affaddr{Hekla, Iceland}\\
%       \email{larst@affiliation.org}
%\and  % use '\and' if you need 'another row' of author names
%% 4th. author
%\alignauthor Lawrence P. Leipuner\\
%       \affaddr{Brookhaven Laboratories}\\
%       \affaddr{Brookhaven National Lab}\\
%       \affaddr{P.O. Box 5000}\\
%       \email{lleipuner@researchlabs.org}
%% 5th. author
%\alignauthor Sean Fogarty\\
%       \affaddr{NASA Ames Research Center}\\
%       \affaddr{Moffett Field}\\
%       \affaddr{California 94035}\\
%       \email{fogartys@amesres.org}
%% 6th. author
%\alignauthor Charles Palmer\\
%       \affaddr{Palmer Research Laboratories}\\
%       \affaddr{8600 Datapoint Drive}\\
%       \affaddr{San Antonio, Texas 78229}\\
%       \email{cpalmer@prl.com}
}
% There's nothing stopping you putting the seventh, eighth, etc.
% author on the opening page (as the 'third row') but we ask,
% for aesthetic reasons that you place these 'additional authors'
% in the \additional authors block, viz.
%\additionalauthors{Additional authors: John Smith (The Th{\o}rv{\"a}ld Group,
%email: {\texttt{jsmith@affiliation.org}}) and Julius P.~Kumquat
%(The Kumquat Consortium, email: {\texttt{jpkumquat@consortium.net}}).}
% Just remember to make sure that the TOTAL number of authors
% is the number that will appear on the first page PLUS the
% number that will appear in the \additionalauthors section.

\maketitle
\begin{abstract}
	Similarity search on time series is a frequent operation in large-scale data-driven applications. Sophisticated similarity measures are standard for time series matching,  as they are usually misaligned. Dynamic Time Warping or DTW is the most widely used similarity measure for time series because it combines alignment and matching at the same time. However, the alignment makes DTW slow. To speed up the expensive similarity search with DTW, branch and bound based pruning strategies are adopted. However, branch and bound based pruning are only useful for very short queries (low dimensional time series), and the bounds are quite weak for longer queries. Due to the loose bounds branch and bound pruning strategy boils down to a brute-force search.
	
	To circumvent this issue, we design SSH (Sketch, Shingle, \& Hashing), an efficient and approximate hashing scheme which is much faster than the state-of-the-art branch and bound searching technique: the UCR suite. SSH uses a novel combination of sketching, shingling and hashing techniques to produce (probabilistic) indexes which align (near perfectly) with DTW similarity measure. The generated indexes are then used to create hash buckets for sub-linear search. Our results show that SSH is very effective for longer time sequence and prunes around 95\% candidates, leading to the massive speedup in search with DTW. Empirical results on two large-scale benchmark time series data show that our proposed method can be around 20 times faster than the state-of-the-art package (UCR suite) without any significant loss in accuracy.
\end{abstract}

%\begin{CCSXML}
%	<ccs2012>
%	<concept>
%	<concept_id>10002951.10003317.10003338.10003346</concept_id>
%	<concept_desc>Information systems~Top-k retrieval in databases</concept_desc>
%	<concept_significance>300</concept_significance>
%	</concept>
%	<concept>
%	<concept_id>10002951.10003317.10003347.10003352</concept_id>
%	<concept_desc>Information systems~Information extraction</concept_desc>
%	<concept_significance>300</concept_significance>
%	</concept>
%	</ccs2012>
%\end{CCSXML}
%
%\ccsdesc[300]{Information systems~Top-k retrieval in databases}
%\ccsdesc[300]{Information systems~Information extraction}


%
%  Use this command to print the description
%
%\printccsdesc

% We no longer use \terms command

%\keywords{Time Series Indexing; Dynamic Time Warping}

\section{Introduction}
Mining for similar or co-related time series is ubiquitous, and one of the most frequent operations, in data driven applications including robotics,medicine~\cite{oates2000method,caracca2000discovering}, speech~\cite{rabiner1993fundamentals}, object detection in vision~\cite{yang2002detecting, sonka2014image}, High Performance Computing (HPC) and system failure diagnosis~\cite{luo2014correlating,sun2014querying}, earth science \cite{mudelsee2013climate}, finance \cite{granger2014forecasting}, and information retrieval \cite{rao2016compressing} etc.

The focus of this paper is on the problem of similarity search with time series data. A time series $X$ is defined as a sequence of values $X = \{x_1,x_2,...,x_m\}$ associated with timestamps:$\{t(x_1), t(x_2),..., t(x_m)\}$ that typically satisfy the relationship $t(x_i) = t(x_{i-1})+\tau$, where $\tau$ is the sampling interval and $m$ is the number of points in the time series. Formally, given a dataset $D = \{X_i|1 \le i \le N\}$ and a query time series $Q$, we are interested in efficiently computing
\begin{equation}
X^* = \arg \max_{X \in D} S(Q,X),
\end{equation}
where $S(X,Y)$ is some similarity of interest between time series $X$ and $Y$. This problem is generally prohibitively expensive for large-scale datasets, especially for latency critical application. We shall concentrate on the computational requirement of this problem.

Finding the right similarity measure for time series is a well-studied problem~\cite{rakthanmanon2012searching}, and the choice of this measure is dependent on the application. It is further well known that while matching time series, it is imperative, for most applications, to first align them before computing the similarity score.  Dynamic time warping or DTW is widely accepted as the best similarity measure (or the default measure) over time series, as pointed out in~\cite{rakthanmanon2012searching}.  DTW, unlike $L_1$ or $L_2$ distances, takes into account the relative alignment of the time series (see Section~\ref{DTWbest} for details). However, since alignment is computationally expensive, DTW is known to be slow~\cite{rakthanmanon2012searching}.

{
%\color{red}

%The significance of the problem combined with the lack of fast hashing schemes leads to a flurry of work which tries to make similarity search with DTW efficient~\cite{ding2008querying,kim2001index,keogh2009supporting} using the branch-and-bound technique.
Owing to the significance of the problem there are flurry of works which try to make similarity search with DTW efficient. The popular line of work use the branch-and-bound technique~\cite{ding2008querying,kim2001index,keogh2009supporting}. Branch and bound methods use bounding strategies to prune less promising candidates early, leading to savings in computations.  A notable among them is the recently proposed UCR suite~\cite{rakthanmanon2012searching}. The UCR suite showed that carefully combining different branch-and-bound ideas leads to a significantly faster algorithm for searching. They showed some very impressive speedups, especially when the query time series is small. UCR suite is currently the fastest package for searching time series with DTW measure, and it will serve as our primary baseline.

Branch and bounds techniques prune down candidates significantly while dealing with small queries (small subsequence search). For short queries, a cheap lower bound is sufficient to prune the search space significantly leading to impressive speedups. However,  when the query length grows, which is usually the case, the bounds are very loose, and they do not result in any effective pruning. Our empirical finding suggests that existing branch-and-bound leads to almost no pruning (less than 1\%, see Section~\ref{issue}) when querying with longer time series, making the UCR suite expensive. % (as shown in Section~\ref{issue}). 
Branch-and-bound techniques, in general, do not scale well when dealing with long time series. Nevertheless, it should be noted that branch and bound techniques give exact answers. It appears that if we want to solve the search problem exactly, just like classical near neighbor search, there is less hope to improve the UCR suite. We will discuss this in details in Section~\ref{issue}.

Indexing algorithms based on hashing are well studied for reducing the query complexity of high-dimensional similarity search \cite{pauleve2010locality,shrivastava2015asymmetric,Proc:Shrivastava_NIPS14}. Hashing techniques are broadly divided into two categories: 1) Data Independent Hashing \cite{Proc:Shrivastava_NIPS14,shrivastava2015asymmetric} and 2) Learning-based (Data Dependent) Hashing \cite{zhang2011composite,zhang2010self}.

To the best of our knowledge, there is only one recent hashing algorithm tailored for the DTW measure: ~\cite{kale2014examination}. This algorithm falls into the category of learning-based hashing.  Here, the authors demonstrated the benefit of kernel-based hashing (Learning-based) scheme~\cite{kulis2009kernelized} for DTW measure on medium scale datasets (60k time series or less). However, the computation of that algorithm scales poorly $O(n^2)$ where $n$ is the number of time series. This poor scaling is due to the kernel matrix ($n \times n$) and its decomposition which is not suitable for large-scale datasets like the ones used in this paper with around 20 million time series.

In addition, the method in \cite{kale2014examination}, as a learning based hashing, requires an expensive optimization to learn the hash functions on data samples followed by hash table construction. Any change in data distribution needs to re-optimize the hash function and repopulate the hash tables from scratch. This static nature of learning-based hashing is prohibitive in current big-data processing systems where drift and volatility are frequent. Furthermore, the optimization itself requires quadratic $O(n^2)$ memory and computations, making them infeasible to train on large datasets (such as the one used in this paper where $n$ runs into millions).

In contrast, data independent hashing enjoys some of the unique advantages over learning-based hashing techniques.
Data independent hashing techniques derive from the rich theory of Locality Sensitive Hashing (LSH)~\cite{gionis1999similarity} and are free from all the computational burden. Furthermore, they are ideal for high-speed data mining in a volatile environment because drift in distribution does not require any change in the algorithms and the data structures can be updated dynamically. Owing to these unique advantages data independent hashing is some of the heavily adopted routines in commercial search engines~\cite{Proc:Henzinger_06}

However, data independent methodologies for time series are limited to vector based distance measures such as $L_p$~\cite{Proc:Datar_SCG04,agrawal1993efficient} or cosine similarity. As argued before, vector based distances are not suitable for time series similarity. Unfortunately, there is no known data independent hashing scheme tailored for the DTW measure. Lack of any such scheme makes hashing methods less attractive for time series mining, particularly when alignments are critical. A major hurdle is to design an indexing mechanism which is immune to misalignments. In particular, the hashes should be invariant to spurious transformations on time series such as shifting. In this work, we provide a data independent hashing scheme which respects alignments, and our empirical results show that it correlates near perfectly with the desired DTW measure. The focus of this paper will be on data-independent hashing schemes which scale favorably and cater the needs of frequently changing modern data distributions.

}

{\bf Our Contributions:} We take the route of randomized hashing based indexing to prune the candidates more efficiently compared to branch-and-bound methods. We propose the first data-independent Hashing Algorithm for Time Series: SSH (Sketch, Shingle, \& Hash).
Indexing using SSH can be around 20x faster than the current fastest package for searching time series with DTW, UCR suite~\cite{rakthanmanon2012searching}. Our proposal is a novel hashing scheme which, unlike existing schemes, does both the alignment and matching at the same time. Our proposal keeps a sliding window of random filters to extract noisy local bit-profiles (sketches) from the time series. Higher order shingles (or $n$-grams with large $n$ like 15 or more) from these bit-profiles are used to construct a weighted set which is finally indexed using standard weighted minwise hashing which is a standard locality sensitive hashing (LSH) scheme for weighted sets.

Our experiments show that the ranking under SSH aligns near perfectly with the DTW ranking. With SSH based indexing we can obtain more than 90\% pruning even with long queries where branch-and-bound fails to prune more than 7\%. Our proposed method is simple to implement and generates indexes (or hashes) in one pass over the time series vector. Experimental results on two large datasets, with more than 20 million time series, demonstrate that our method is significantly (around 20 times) faster than the state-of-the-art package without any noticeable loss in the accuracy.

The rest of this paper is organized as follows. Section \ref{background} introduces background of our work. In Section \ref{issue} we discuss why pruning strategies can not work well when dealing with long queries.
We then describe our approach in Section \ref{algorithm}. Section \ref{experiment} presents our experimental results.

\section{Background}
\label{background}%
Let us now review several backgrounds of our work. We introduce DTW (Dynamic Time Warping) for time series in Section \ref{DTWbest}. We then introduce Locality Sensitive Hashing and Weighted Minwise Hashing in Section \ref{sec:lsh}.
%In this section, we introduce notations and definitions used in this paper and claim that DTW is the best similarity measure. The then introduce the related works of similarity search in time series.
%
%\subsection{Notation and Definition}
%
%In this section, we introduce some notions of time series.
%
%A time series is defined as follow:
%\begin{definition}[Time Series]
%A time series, denoted as $X = (x_1,x_2,...,x_m)$, where $m$ is the number of points in the time series. The timestamps of a time series, denoted as $TX = (t(x1), t(x2),..., t(xn))$, have the relationship of $t(x_i) = t(x_{i-1})+\tau$, where $\tau$is the sampling interval.
%\end{definition}
%
%In this paper, we interested in the problem of time series similarity search problem:
%\begin{definition}[Top $1$ Similarity Search]
%Given a data set $D = \{X_i|0<i<N-1\}$, where $N$ denotes the size of this data set. Given a query time series $X_q$, the top $1$ similarity search is to find the a time series $X^* \in D$, where
%
%\begin{equation}
%\label{knn}
%X^* = \arg \max_{X \in D} S(X_q,X)
%\end{equation}
%
%
%where $S(X,Y)$ denote the similarity between time series $X$ and $Y$. In this work, we are more interested in the problem of top-k similarity search problem, which is finding the top $k$ most similar time series to $X_q$ in $D$. It is pointed out that, the size $N$ of data set $D$ often huge (i.e. million level). And in this paper, our aim is to do fast searching top $k$ time series.
%\end{definition}

\subsection{Dynamic Time Warping and Expensive Computation}
\label{DTWbest}

\begin{figure*}[t]
	\centering
	\includegraphics[width=4.5in]{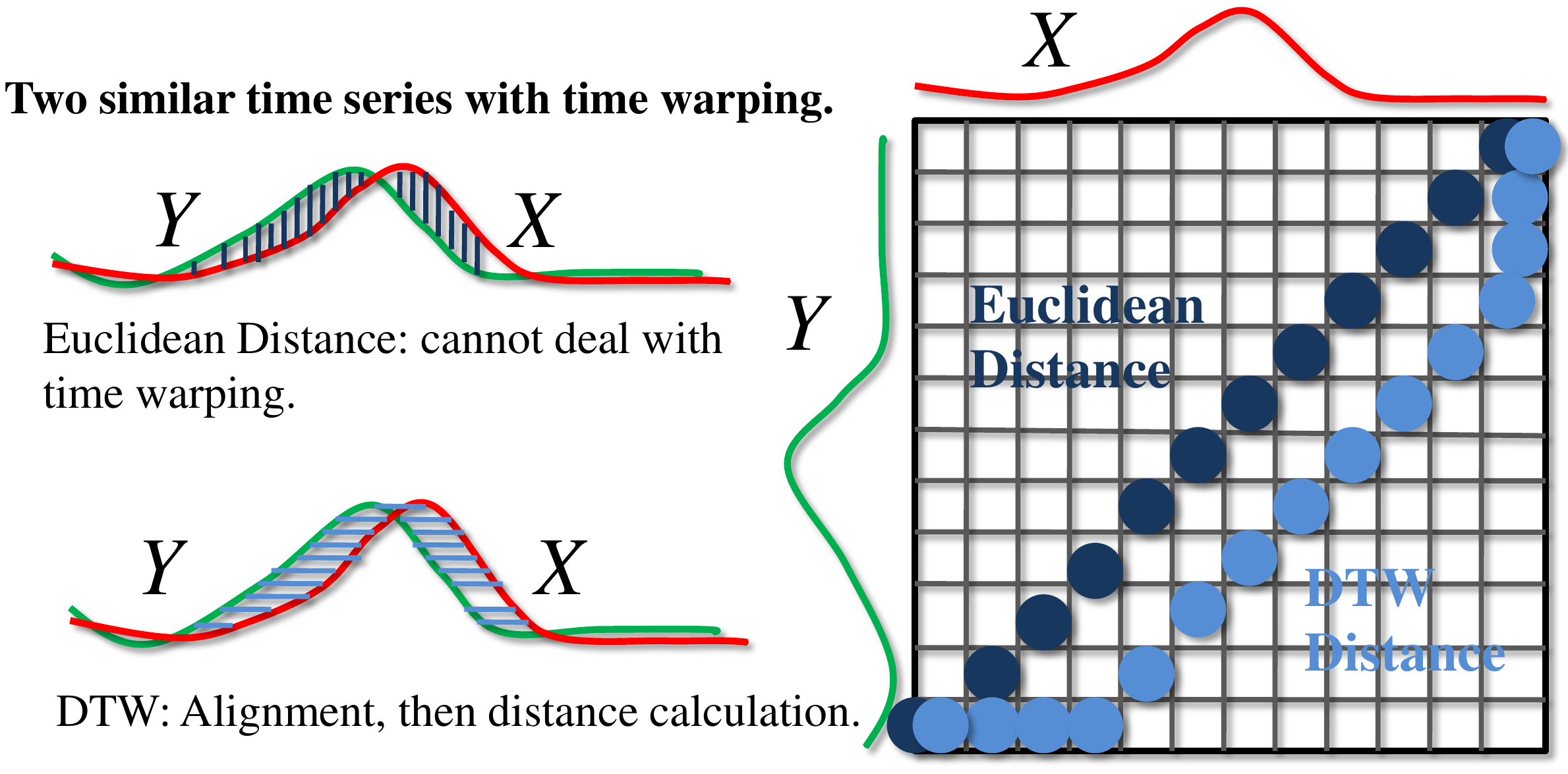}
	\caption{The difference between Euclidean and DTW distances of two time series $X$ and $Y$. The DTW distance computes the similarity of the best alignment and hence can deal with time warping of $X$ and $Y$.}
	\label{fig:dtw}
\end{figure*}

%One of the problems that effect the time series similarity search is the choosing of the similarity measure.
One of the peculiarities of time series similarity which is different from general vector similarity is its invariance with warping or shift in time. For example, a series $X = \{x_1,x_2,...,x_m\}$, associated with timestamps:
\[\{t(x_1), t(x_2),..., t(x_m)\}\]
should be very similar to a slightly shifted time series $X' = \{x_3,x_4,....,x_m,y,z\}$ over the same time stamps. This high similarity is because there is a significantly long subsequence of $X$ and $X'$, which are identical (or very similar). Traditional measures such as $L_2$ distance are not suitable for such notions of similarity as they are sensitive to shifts. Dynamic Time Warping (DTW) was designed to align various systematic inconsistencies in the time series, which is the main reason behind its wide adoption.

To compute the DTW distance we construct an $m$-by-$m$ matrix $W$, where the ($i$-th,$j$-th) element of the matrix $W$ denotes the difference between $i$-th component of $X$ and $j$-th component of $Y$. The DTW distance finds the path through the matrix that minimizes the total cumulative distance between $X$ and $Y$ (Fig. \ref{fig:dtw}).
The optimal path is the one that minimizes the warping cost:
\[DTW(X,Y) = \min {\sqrt{\sum_{k=1}^{K}w_k}}\]
where, $w_k$ is the $k-{th}$ element of a warping path $P$, which is a contiguous set of elements that represent a mapping between $X$ and $Y$. The overall computation of DTW is given by a dynamic program, please see~\cite{rakthanmanon2012searching} for more details.

DTW is costly as it requires $O(m^2)$ computations using a dynamic programming solution, where $m$ is the time series length. DTW computes the optimal alignment of the two given time series followed by calculating the optimal similarity after the alignment. As expected, alignment is a slow operation. To make searching, with DTW, efficient a common strategy is to resort of branch and bound based early pruning~\cite{rakthanmanon2012searching}.

\subsection{Locality Sensitive Hashing and Weighted Minwise Hashing}
\label{sec:lsh}
\subsection{Locality Sensitive Hashing (LSH) }

Locality-sensitive hashing (LSH)~\cite{andoni2006near,li2011hashing} is common for sub-linear time near neighbor search. The basic idea of LSH is to hash input items to different buckets so that similar items map to the same ``buckets" with high probability.

LSH generates a random hash map $h$ which takes the input (usually the data vector) and outputs a discrete (random) number. For two data vectors $x$ and $y$, the event $h(x) = h(y)$ is called the collision (or agreement) of hash values between $x$ and $y$.
The hash map has the property that similar data vectors, in some desired notion, have a higher probability of collisions than non-similar data vectors. Informally, if $x$ and $y$ are similar, then $h(x) = h(y)$ is a more likely event, while if they are not similar then $h(x) \ne h(y)$ is more likely. The output of the hash functions is a noisy random fingerprint of the data vector~\cite{Proc:Carter_STOC77,Report:Rabin_81,karp1987efficient}, which being discrete is used for indexing training data vectors into hash tables. These hash tables represent an efficient data structure for similarity search~\cite{Proc:Indyk_STOC98}.

For the details of Locality-sensitive hashing, please refer~\cite{andoni2006near,li2011hashing}.

\subsection{Weighted Minwise Hashing}

Weighted Minwise Hashing is a known LSH for the Weighted Jaccard similarity~\cite{leskovec2014mining}. Given two positive vectors $x, \ y \in \mathbb{R^D}$, $x, \ y > 0$, the (generalized) Weighted Jaccard similarity is defined as
\begin{equation}\label{eq:WJaccardDef}
\mathbb{J}(x,y) = \frac{\sum_{i=1}^D\min\{x_i,y_i\}}{\sum_{i=1}^D\max\{x_i,y_i\}}.
\end{equation}
$\mathbb{J}(x,y)$ is a frequently used measure for comparing web-documents~\cite{Proc:Broder}, histograms (specially images), gene sequences, etc. Recently, it was shown to be a very effective kernel for large-scale non-linear learning~\cite{Proc:Li_KDD15}. WMH leads to the best-known LSH for $L_1$ distance, commonly used in computer vision, improving over~\cite{Proc:Datar_SCG04}.

Weighted Minwise Hashing (WMH) (or Minwise Sampling) generates randomized hash (or fingerprint) $h(x)$, of the given data vector $x \ge 0$,  such that for any pair of vectors $x$ and $y$, the probability of hash collision (or agreement of hash values) is given by,
\begin{equation}\label{eq:CollProb}
Pr(h(x) = h(y)) = \frac{\sum \min\{x_i,y_i\}}{\sum \max\{x_i,y_i\}} = \mathbb{J}(x,y).\end{equation} A notable special case is when $x$ and $y$ are binary (or sets), i.e. $x_i, y_i \in \{0,\ 1\}^D$ . For this case, the similarity measure boils down to $\mathbb{J}(x,y) = \frac{\sum \min\{x_i,y_i\}}{\sum \max\{x_i,y_i\}} = \frac{|x \cap y|}{|x \cup y|}$.

Weighted Minwise Hashing (or Sampling), \cite{Proc:Broder,Proc:Broder_WWW97,Report:Manasse_00} is the most popular and fruitful hashing technique for indexing weighted sets, commonly deployed in commercial big-data systems for reducing the computational requirements of many large-scale search~\cite{Proc:Broder_FUN98,Proc:Bayardo_WWW07,Article:Henzinger04,Proc:Henzinger_06,Proc:Koudas_SIGMOD06,Proc:Chien_WWW05}.
Recently there has been many efficient methodologies to compute weighted minwise hashing~\cite{Report:Manasse_00,ioffe2010improved,Proc:OneHashLSH_ICML14,Proc:Shrivastava_UAI14,Report:Haeupler_arXiv14,Proc:Shrivastava_NIPS16}.

\section{Longer Subsequences and Issues with Branch and Bound}
\label{issue}

\begin{table*}
	\caption{Percentage of candidates that pruned by UCR Suite on ECG Data set and Random Walk Data set. With the increasing of the time series length, the ability of lower bounds used by UCR Suite to prune candidates deteriorate as the bounds suffer from the curse of dimensionality.}
	\label{pruning}
	\centering
	\begin{tabular}{lllll}
		\toprule
		Time Series Length     & 128   & 512  & 1024  & 2048\\
		%		\midrule
		%		LB\_Kim & 70.63\%  & 0.03\% & 0.01\% & 0.00\%     \\
		%		LB\_Keogh & 29.34\%  & 93.46\% & 8.65\% & 0.88\%    \\
		%		LB\_Keogh2 & 2.06\%  & 1.47\% & 10.04\% & 6.88\%    \\
		%		\hline
		\hline
		UCR Suite Pruned (ECG)  & \textbf{99.7\%} & \textbf{94.96\%} & \textbf{18.70\%} & \textbf{7.76\%} \\
		UCR Suite Pruned (Random Walk)& \textbf{98.6\%} & \textbf{14.11\%} & \textbf{30.2\%} & \textbf{3.5\%} \\
		\bottomrule
	\end{tabular}
\end{table*}

Branch and bound strategies are used for reducing the searching cost by pruning off bad candidates early. The core idea behind branch and bound is to keep a cheap-to-compute lower bound on the DTW distance. For a given query, if the lower bound of the current candidate exceeds the best seen DTW then we ignore this candidate safely, simply using cheap lower bounds. This strategy eliminates the need for computing the costly DTW.

UCR Suite \cite{rakthanmanon2012searching} combines several branch and bound strategies and makes time series searching process very fast.  Three main branch and bound strategies are used in UCR Suite \cite{rakthanmanon2012searching}: $LB_{Kim}$ \cite{kim2001index} lower bound, $LB_{Keogh}$ lower bound, and $LB_{Keogh2}$ lower bound \cite{keogh2009supporting}.
$LB_{Kim}$ uses the distance between the First (Last) pair of points from Candidate time series and the Query time series as the lower bound. The complexity of calculating $LB_{Kim}$ is $O(1)$.
$LB_{Keogh}$ and $LB_{Keogh2}$ \cite{keogh2009supporting} uses the Euclidean distance between the candidate series and Query series.

The complexity of this lower bound is $O(n)$, where $n$ is the time series length. These three branching and bounds strategies can prune bad candidates in $O(1)$ (or O(n)) time which are significantly smaller than the time needed to compute DTW distance ($O(n^2)$ time).

{
%\color{red}

However, the lower bound gets weaker with the increase in the length of the time series, due to the curse of dimensionality. This weakening of bounds with dimensionality makes branch-and-bound ideas ineffective. We demonstrate this phenomenon empirically on two large-scale datasets (also used in our experiment see Section \ref{experiment}). Table.\ref{pruning} shows the percentage of candidates that are pruned by the three lower bounding strategies as well as the UCR Suite which combines all the three.

The code of this experiment are taken from the UCR Suite package \footnote{http://www.cs.ucr.edu/~eamonn/UCRsuite.html}, which is publicly available. This implementation of UCR Suite uses three pruning lower bound: $LB_{Kim}$ \cite{kim2001index} lower bound, $LB_{Keogh}$ lower bound, and $LB_{Keogh2}$ lower bound \cite{keogh2009supporting}. For each time series, this UCR Suite package compares all the three lower bound for each time series and uses the lowest one to do pruning.

From Table.\ref{pruning} we can see that when the time series is short (e.g. $128$), the pruning strategies performs quite well ($98\%$ to $99\%$ of the time series pruned). However, when the time series length is around 1000 or more, then all the three criteria are completely ineffective (only $3\%$ to $7\%$ of the time series pruned), and the search boils down to nearly brute force.
We observe the same trend on both the datasets as evident from Table. \ref{pruning}.

Intuitively, as the length of the query time series increases, the number of possible good alignments (or warping) also increases. A myopic $O(n)$ lower bound is, therefore, not effective to eliminate all the possibilities.

%The above example shows that branch and bound strategies can not do efficient pruning for longer time series.
%As a result, in this paper, we propose the SSH (Sketch, Shingle, and Hash) framework for indexing the time series data, which can lead to $95\%$ pruning even for longer time series during searching. We will introduce in details in Section \ref{algorithm}.

}

%
%\begin{table}
%	\caption{Percentage of candidates that pruned by UCR Suite, including three classical  branch-and-bound strategies, on Randomwalk Data set. With the increasing of the time series length, the ability of these lower bounds to prune candidates deteriorate as the bounds suffer from the curse of dimensionality.}
%	\label{pruning_rand}
%	\centering
%	\begin{tabular}{lllll}
%		\toprule
%		Time Series Length     & 128   & 512  & 1024  & 2048\\
%		\midrule
%%		LB\_Kim & 60.93\%  & 5.13\% & 0.61\% & 0.01\%     \\
%%		LB\_Keogh & 22.58\%  & 63.46\% & 7.85\% & 1.28\%    \\
%%		LB\_Keogh2 & 15.09\%  & 70.06\% & 21.74\% & 2.21\%    \\
%%		\hline
%		UCR Suite Pruned & \textbf{98.6\%} & \textbf{14.11\%} & \textbf{30.2\%} & \textbf{3.5\%} \\
%		\bottomrule
%	\end{tabular}
%\end{table}

\section{Our Proposal: SSH (Sketch, Shingle \& Hash)}
\label{algorithm}

%We are  the problem of top-$k$ similarity search, the input of LSHTS, is a data set $D = \{X_i|0<i<N-1\}$, and a query $Q$.
{\bf SSH (Sketch, Shingle \& Hash):} We propose a new hashing scheme, for time series, such that hash collisions are ``active" indicator of high similarity while ignoring misalignments if any. Our hashing scheme consists of the following stages:
\begin{enumerate}
	\item {\bf Sliding Window Bit-profile (Sketch) Extraction:} We use sliding window of random filter to generate a binary string $B_X$ (sketch) of the given time series $X$.
	\item {\bf Shingle(n-grams) Generation:} We generate higher order shingles from the bit string $B_X$. This process generates a weighted set $S_X$ of shingles.
	\item {\bf Weighted MinHash Computation:} Our final hash value is simply the weighted minwise hashes of $S_X$, which we use as our indexes to create hash tables for time series.
\end{enumerate}

{
Next, we go over each of the three steps in detail.
}
\subsection{Sliding Window Bit-profile Extraction}
\label{InforExtract}

\begin{figure*}[t]
	\centering
	\includegraphics[width=0.8\textwidth]{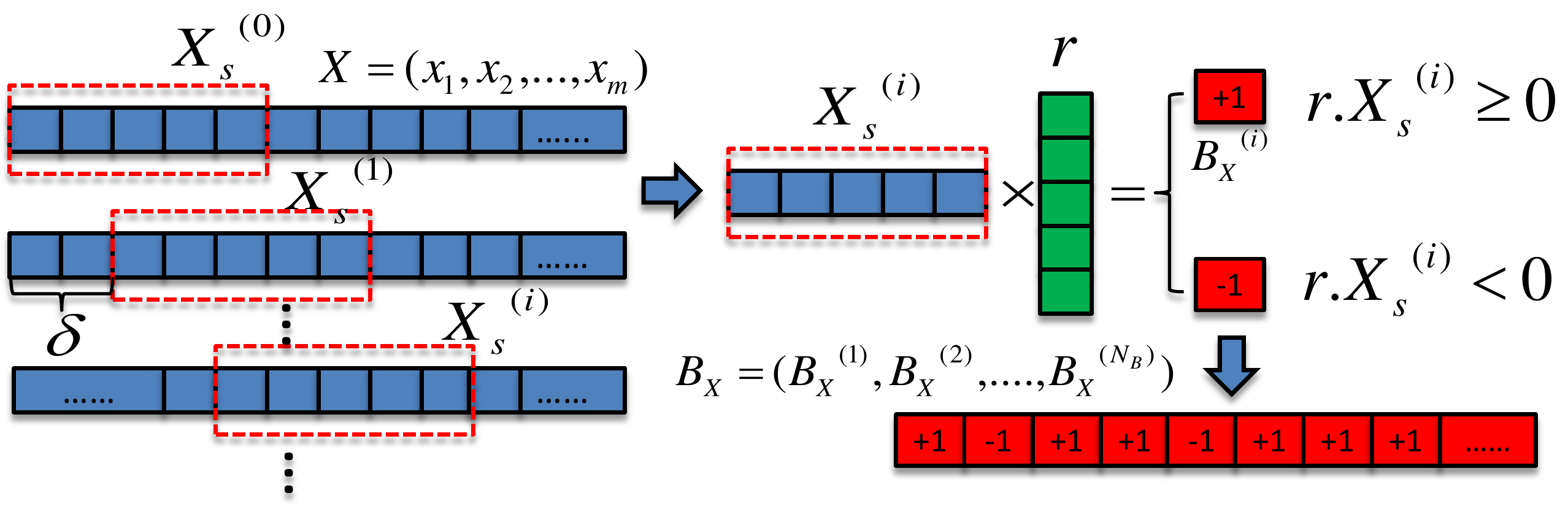}
	\caption{For each time series $X$, we convolve it with a sliding window (red dash box), with shift $\delta$, of random gaussian filter $r$ and generate a bit depending on the sign of the inner product. After the complete slide, the process generate a binary string (sketch) $B_X$ which captures the pattern in the time series.}
	\label{fig:sliding}
\end{figure*}

We have a successful set of methodologies based on shingling~\cite{leskovec2014mining} to deal with massive-scale discrete sequential data such as text or strings.  However, time series data contains continuous values making shingling based approaches inapplicable.  Furthermore, variations in sampling intervals, frequency, and alignments make the problem worse.

Our first step solves all this problem by converting time series with continuous values to discrete sequential objects which can be handled using shingling later.   We use the idea of sketching time series with a sliding window of random filters~\cite{Indyk:2000:IRT:645926.671699}, which was shown to capture trends effectively.  In particular, we produce a bit string (sketch) from the time series. Each bit in this string captures crude information of some small subsequence in the time series.

To generate local bit-profile, we use a randomly generated filter which is a small vector $r$, of appropriately chosen size $W$, as shown in Figure~\ref{fig:sliding}. This filter slides over the time series with an appropriately selected step size $\delta$. During each slide, the filter $r$ is multiplied to the current $W$ length subsequence of the time series, and a bit indicating the sign of the output is stored.  In technical terms, this is a signed projection of the selected window~\cite{Article:Li_Konig_CACM11,Proc:Charikar}. These crude sketches are robust to various perturbations in the values of time series.

More formally, Given a time series $X = (x_1,x_2,...,x_m)$,  the length $W$ of vector $r$, step size $\delta$. The extracted information is a (bit) sign stream, given by:
\begin{equation}
B_X = (B_X^{(1)},B_X^{(2)},...,B_X^{(N_B)}).
\end{equation}
Where $N_B = (m-W)/\delta$ is the size of the sign stream $B_X$. And each $B_X^{(i)}$s is calculated as follow:
\begin{equation}
B_X^{(i)} = \left\{\begin{matrix}
+1~&:~ r.X_s^{(i)} \geq 0 \\
-1~&:~ r.X_s^{(i)} < 0
\end{matrix}\right.
\end{equation}
In above, $X_s^{(i)} = \{ x_{i*\delta}, x_{i*\delta + 1}, ... , x_{i*\delta + W - 1} \}$ is the sub-series of length $W$.

\label{ngram}

For example, given a time series $X=(1,2,4,1)$, a small filter $r=(0.1,-0.1)$, and a step size $\delta=2$. Then the extracted sign stream is:

\begin{equation} \label{eq1}
\begin{split}
B_X & = (sign((1,2)*(0.1,-0.1)),sign((4,1)*(0.1,-0.1))) \\
& = (sign(-0.1),sign(0.3))\\
& = (-1,+1)
\end{split}
\end{equation}

In this step, we choose $r$ as a spherically symmetric random vector with length $W$, i.e. the entries of $r$ are $i.i.d$ normal, i.e., $r \sim N(0,1)$. This choice is a known locality sensitive hashing for cosine similarity~\cite{li2011hashing}. From the theory of signed random projections~\cite{li2011hashing} these bits are 1-bit dimensionality reduction of the associated small subsequence which was multiplied by $r$. It ensures that bit matches are a crude probabilistic indicator of the closeness of local profiles.

\subsection{Shingle(n-grams) Generation}

After the sketching step we have a bit-string profile $B_X$ from the time series $X$, where the $i^{th}$ bit value $B_X(i)$ is a 1-bit summary (representation) of a small subsequence of X, i.e. $X_s^{(i)} = \{ x_{i*\delta}, x_{i*\delta + 1}, ... , x_{i*\delta + W - 1} \}$. Therefore, for two vectors $X$ and $Y$, $B_X(i) = B_X(j)$ indicates that the small subsequences $X_s^{(i)} = \{ x_{i*\delta}, x_{i*\delta + 1}, ... , x_{i*\delta + W - 1} \}$ and $Y_s^{(j)} = \{ y_{j*\delta}, y_{j*\delta + 1}, ... , y_{i*\delta + W - 1} \}$ are likely to be similar due to the LSH property of the bits.

\subsubsection{Intuition of why this captures alignments as well as similarity?}
\begin{figure}[t]
	\centering
	\includegraphics[width=0.5\textwidth]{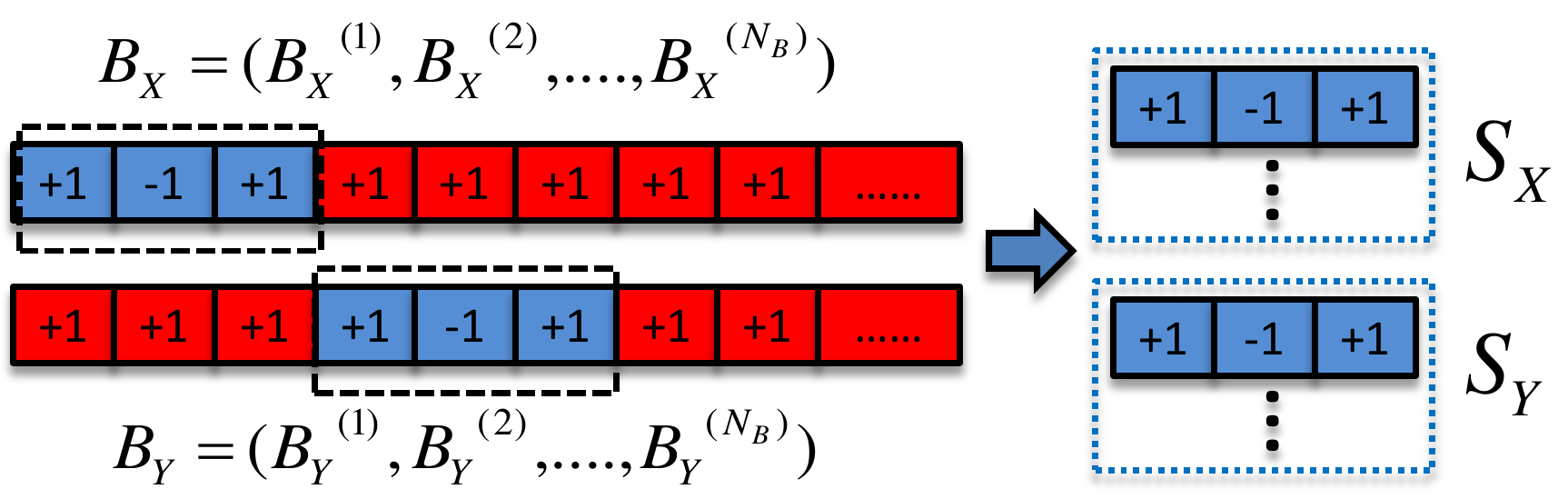}
	\caption{{\bf SSH Illustration on two different time series:} Two different time series $X$ and $Y$ has same pattern (Blue window). We use n-gram to extract patterns and use the pattern set $S_X$ and $S_Y$ to represent the time series, then the time warping of time series is solved by set similarity.}
	\label{ngramalign}
\end{figure}

If for two time series $X$ and $Y$ there is a common (or very similar) long subsequence, then we can expect that a relatively large substring of $B_X$ will match with some other significant substring of $B_Y$ (with possibly some shift). However, the match will not be exact due to the probabilistic nature of bits. This situation is very similar to the problem of string matching based on edit distance, where token based (or n-gram based) approach has shown significant success in practice. The underlying idea is that if two bit strings $B_X$ and $B_Y$ has a long common (with some corruption due to probabilistic nature) subsequence, then we can expect a significant common $n$-grams (and their frequency) between these bit strings. It should be noted that n-gram based approach automatically takes care of the alignment as shown in Fig. \ref{ngramalign}.

For example, given a bit string $B_X=(+1,+1,-1，-1，+1，+1)$, and shingle length $n=2$, then we can get a weighted set:
\[
S_X = \{(+1,+1):2, (+1,-1):1, (-1,+1):1, (-1,-1):1\}.
\]

\begin{figure}[t]
	\centering
	\includegraphics[width=0.5\textwidth]{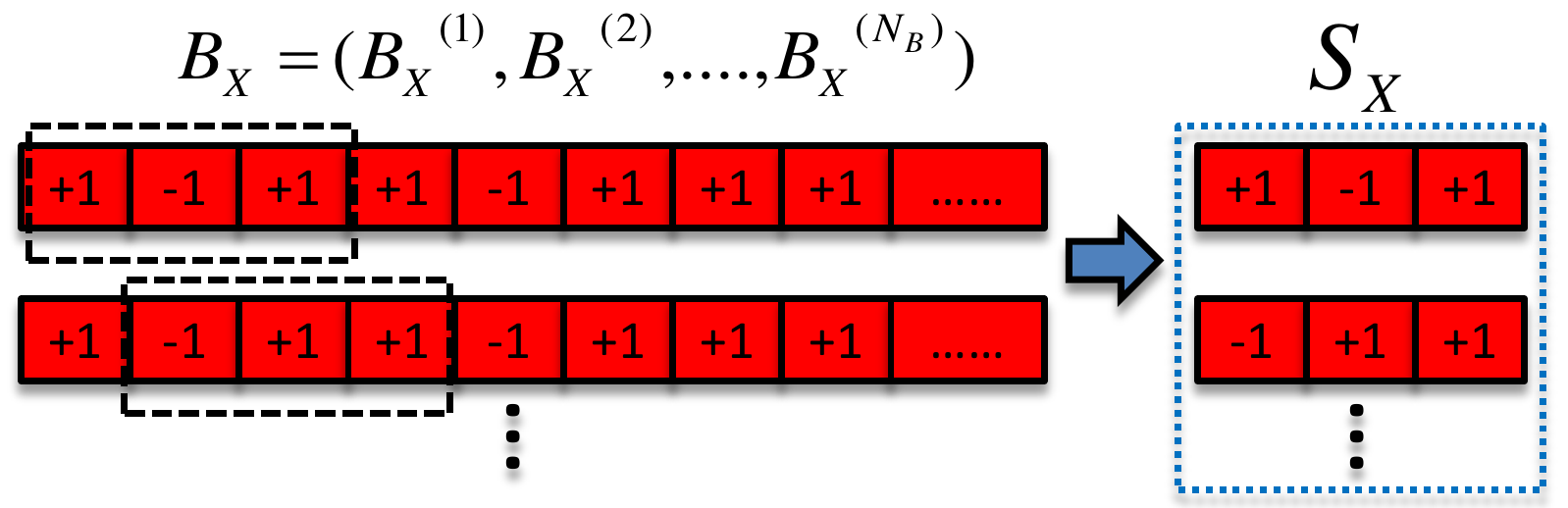}
	\caption{Shingle(n-grams) Generation: Give the bit string sketch generated from step-1, we treat it as string and generate n-grams shingles. The shingling process outputs a weighted set.}
	\label{ngramconstruct}
\end{figure}

Formally, given the bit (or sign) stream
\[B_X = (B_X^{(1)},B_X^{(2)},...,B_X^{(N_B)})\] for a time series $X$, we construct weighted set $S_{X}$ by including all substrings of length $n$ ($n$-gram) occurring in $B_X$ with their frequencies as their corresponding weight (see Figure.\ref{ngramconstruct}).
\begin{equation}
\label{set}
S_{X} = \{ S_i, w_i~|~S_i = \{ B_X^{(i)},B_X^{(i+1)}, ..., B_X^{(i+n-1)}\} ~,~ 0<i<n \}
\end{equation}
Notice that, the set is a weighted set, $w_i$ denotes the number of tokens (or patterns) $S_i = \{ B_X^{(i)},B_X^{(i+1)}, ..., B_X^{(i+n-1)}\}$ present in the time series. The intuition here is that the Weighted Jaccard similarity between the sets $S_X$ and $S_Y$,  generated by two different time series $X$ and $Y$, captures the closeness of the original time series. This closeness is not affected by spurious shifting of the time series.

\subsection{Weighted MinHash Computation}
\label{sec:weighted}

The Shingle(n-grams) generation step generates a weighted set $S_X$ for the given time series $X$. Since we want to capture set similarity, our final hash value is simply the weighted minwise hashing of this set. We use these weighted minwise hashes as the final indexes of the time series $X$, which can be utilized for creating hash tables for sub-linear search. 

Weighted minwise hashing (or Consistent Weighted Sampling) is a standard technique for indexing weighted sets~\cite{Proc:Broder_WWW97}. There are many efficient methodologies to compte them~\cite{Report:Manasse_00,ioffe2010improved,Proc:OneHashLSH_ICML14,Proc:Shrivastava_UAI14,Report:Haeupler_arXiv14,Proc:Shrivastava_NIPS16}. Please refer to~\cite{Proc:Shrivastava_NIPS16} for details.

\subsection{Overall Framework}

\begin{figure*}[t]
	\centering
	\includegraphics[width=0.8\textwidth]{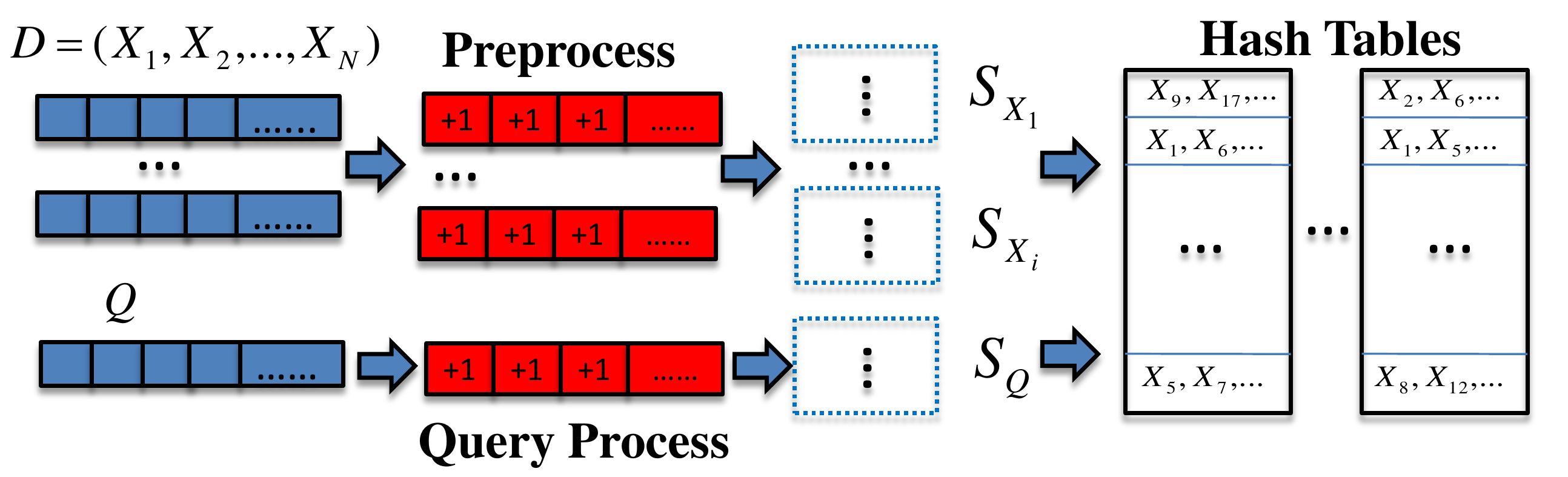}
	\caption{Overall framework contains two steps: (1) Preprocess and (2) Query process. In Preprocess stage, all the time series in data set $D$ hased in to hash tables following three processing steps: (1) Sliding Window Bit Profile Extraction, (2) Shingle (n-gram) Generation, (3) Weighted MinHash Computation. In Query Process, given a query time series, find the associated buckets in the hash tables using the same s-step hashing schema.}
	\label{fig:frame}
\end{figure*}

Given a time series search data set $D = \{X_i|1\le i \le N\}$, the query time series $Q$, and the corresponding parameters $W$, $r$, $\delta$. Our goal is to output the top-$k$ most similar time series of $Q$.
The proposed framework contains two steps (1) Preprocessing Step: preprocess all the time series, and hash them into hash tables using our 3-step SSH scheme (2) Query step, given a time series, find the associated buckets in the hash tables using the same 3-step SSH scheme. Select top-$k$ among the candidates retrieved from the buckets. The detailed steps of our proposed framework are illustrated in Figure.\ref{fig:frame} and summarized in Algorithm \ref{HaTSPre}, and Algorithm \ref{HaTSQue}.

\begin{algorithm}[h]
	\caption{Pre-Processing}
	\label{HaTSPre}
	\begin{algorithmic}[1]
		\State Input: Given $D = \{X_i|0<i<N-1\}$, the sub-series of length $W$. a spherically symmetric random vector $r$ with length $W$, step size $\delta$, n-gram Shingle Length $n$, number of hash tables $d$.
		
		\State Output: Constructed $d$ hash tables.
		
		\State Initialization $i=0$
		\For{Each time series $S_i$ in $D$}
		
		\State Extract the information of time series $S_i$ using the method introduced in Section. \ref{InforExtract}.
		\State Using $n$-gram method introduced in Section \ref{ngram}.
		\State Using weighted minhash algorithm introduced in Section \ref{sec:weighted} to hash each time series into $d$ different hash tables.
		\EndFor
		
		\State \Return Constructed $d$ hash tables.
	\end{algorithmic}
\end{algorithm}

Algorithm \ref{HaTSPre} shows the Preprocessing stage using the SSH scheme. This stage takes the time series data sets $D = \{X_i|1\le i \le N\}$ as input, and construct $d$ hash tables. In Algorithm \ref{HaTSPre}, for each time series $S_i$ in $D$, we perform Sliding Window Bit Profile Extraction (line 5), Shingle (n-gram) Generation (line 6), and Weighted MinHash Computation (line 7). These three SSH steps (line 5-7) hashes the time series into appropriate hash tables for future queries.

%\textbf{Preprocessing Step:}
%\begin{itemize}
%    %\item[~] /*Preprocessing Step*/
%    \item[Step1.] Initialize Hash Tables, $W$, $\delta$, $r$ and random seeds for weighted minhash.
%    \item[Step2.] For Each time series $X_i$ in $D$
%    \begin{itemize}
%        \item[Step2.1] Extract the bit information $B_X$ of time series $X_i$ using Sliding Window Bit-profile Extraction. (Section \ref{InforExtract}.)
%        \item[Step2.2] Generate weighted set $S_X$ using Shingle(n-grams) Generation method. (Section \ref{ngram}).
%        \item[Step2.3] Use $L$ independent weighted minhash over $S_X$ (Section \ref{sec:weighted}) to hash each time series into $L$ different hash tables.
%    \end{itemize}
%\end{itemize}

\begin{algorithm}[h]
	\caption{Query Process}
	\label{HaTSQue}
	\begin{algorithmic}[1]
		\State \textbf{Input:} Given $D = \{X_i|0<i<N-1\}$, the sub-series of length $W$. a spherically symmetric random vector $r$ with length $W$, step size $\delta$, n-gram Shingle Length $n$, and the number of return series $k$.
		
		\State \textbf{Output:} Top $k$ time series in $D$.
		
		\State Extract the information of time series $Q$ using the method introduced in Section. \ref{InforExtract}.
		\State Using $n$-gram method introduced in Section \ref{ngram} to get the weighted set of $Q$.
		\State Using weighted minhash algorithm introduced in Section \ref{sec:weighted} to ge the hash value of $Q$.
		
		\State Initialize the retrieved set $R$ to null
		\For{Each Hash table $T_i$}
		\State Add all the time series in the probed bucket to $R$
		\EndFor
		
		\State \Return Search $R$ for top-$k$ time series using UCR Suite algorithm
	\end{algorithmic}
\end{algorithm}

Algorithm \ref{HaTSQue} shows the querying process with the SSH scheme. This stage takes the query time series $Q$ as input and returns top-$k$ time series. We use the same SSH steps, Sliding Window Bit Profile Extraction (line 3), Shingle (n-gram) Generation (line 4), and Weighted MinHash Computation (line 5) on the query time series $Q$ to generate the indexes.  Using these indexes, we then probe the buckets in respective hash tables for potential candidates.  We then report the top-$k$ similarity time series, based on DTW  (line 7-10).
The reporting step requires full computation of DTW between the query and the potential candidates.  To obtain more speedups, during the last step, we use the UCR suite branch-and-bound algorithm to prune the potential candidate further.

\subsection{Discussions and Practical Issues}
\label{sec:practical}

{%\color{red}

The SSH procedure leads to a weighted set which combines noisy sketching with cheap shingle based representation. Shingling (or Bag-of-Words or n-grams statistics) is a very powerful idea and has led to state-of-the-art representations for a variety of structured data which includes text, images, genomes, etc. It is further known that shingling is a lossy description because it does not capture complete information of the sequence data, and therefore do no have provable guarantees. Nevertheless, reasonably higher order shingles are still the best performing methods in the information retrieval task with both text and image datasets. For example, the state-of-the-art method for image retrieval, as implemented in popular openCV~\cite{bradski2008learning} package,  compute various noisy features such as SIFT and then represent the image as bag-of-words of those SIFT features.  The main argument that goes in favor of noisy representations is that real world high-dimensional datasets come from a distribution which makes the problem much simpler than the combinatorial hardness associated with their raw representations. A noisy representation many times is good enough to capture the essence of that distribution and thus can save significantly over methods which try to solve the problem exactly in the combinatorial formulation.

In SSH procedure there are three main parameters:  Length $W$ of the spherically symmetric random vector $r$, step size $\delta$, and n-gram Shingle Length $n$. Different choice of these parameters will impact the performance of SSH.

%In some cases, the value of $W$ can be selected based on domain knowledge. For example, the best $W$ for a periodic time series data set is just the period length. The value $\delta$ is highly correlated with the locality constraint of the DTW computation. In most of the cases, we can directly assign the value of $\delta$ as the local constraint of DTW computation. For the selection of shingle Length $n$. A smaller $n$ (e.g. $n=1$) can obtain more information, thus produce better indexing result. However, smaller $n$ will lead to a huge amount of computation.
%The trade off of choosing $n$ here depends on the real applications.

As with other shingling methods, the right selection of the parameters is usually dependent on the data and the similarity distribution.  We can easily choose these parameters using a holdout dataset that suits our task. Since we are interested in retrieving with the DTW measure, we determine values of these parameters such that the rankings under hash collisions nearly agree with the ranking of DTW over a small sample of the dataset. We introduce details, and thorough analysis of these parameters study in section \ref{experiment}.

It should be further noted that the overall hashing scheme can be computed in just one pass over the time series. As we scan, can keep a sliding window over the time to calculate the inner product with filter $r$. Each bit generated goes into a buffer of size $n$. For every n-gram generated, we can hash the tokens and update the minimum on the fly.
%\subsubsection{Making The Set Representation More Informative}
}
%In SSH procedure there are three main parameters:  Length $W$ of the spherically symmetric random vector $r$, step size $\delta$, and n-gram Shingle Length $n$. Different choice of these parameters will impact the performance of SSH.
%Since we are interested in retrieving on DTW measure, we can choose these parameters such that the rankings under hash collisions nearly agree with the ranking of DTW over a small sample of the dataset. We will introduce details, and thorough analysis of these parameters study in section \ref{experiment}.
%
%It should be further noted that the overall hashing scheme can be computed in just one pass over the time series. As we scan, can keep a sliding window over the time to calculate the inner product with filter $r$. Each bit generated goes into a buffer of size $n$. For every n-gram generated, we can hash the tokens and update the minimum on the fly.

\section{Experiment}
\label{experiment}

In this section, we describe our experimental evaluation of SSH procedure on two benchmark data set: ECG time series data and Random Walk time series data. Since our proposal is a new indexing measure for DTW similarity, just like~\cite{rakthanmanon2012searching}, our gold standard accuracy will be based on the DTW similarity.

The experiments run on a PC (Xeon(R) E3-1240 v3 @ 3.40GHz $\times$ 8 with 16GB RAM). All code are implemented in C++. We use g++ 4.8.4 compiler. To avoid complications, we do not use any c++ compiler optimization tools to speed up the program.

\begin{figure*}
	\centering
	\includegraphics[width=\textwidth]{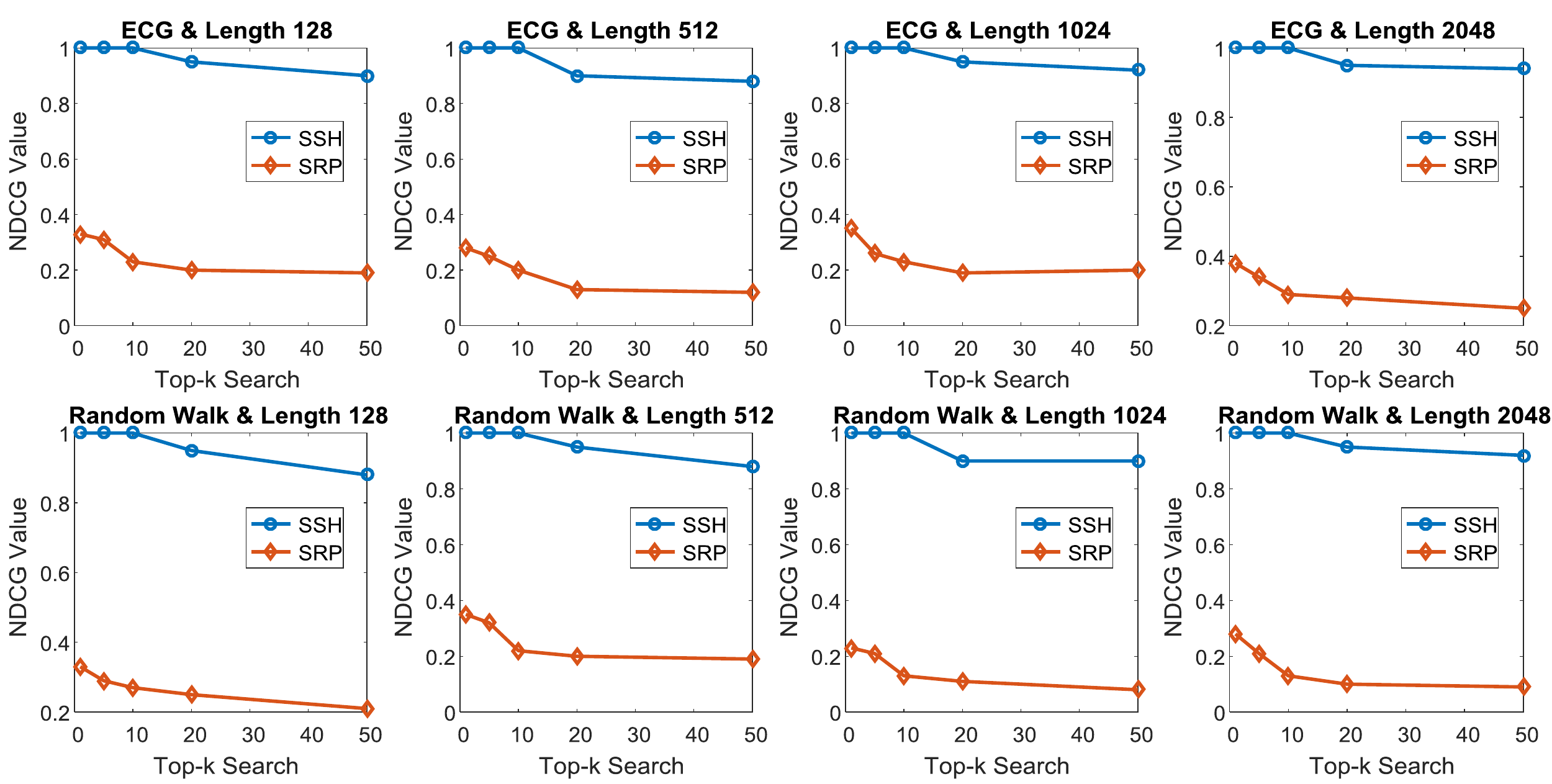}
	\caption{The NDCG (Normalized Discounted Cumulative Gain) of SSH and SRP (Sign Random Projection) on ECG and Random Walk Datasets. The Gold Standard Ranking was based on DTW Distance.}
	\label{fig:ndcg}
\end{figure*}

\begin{table*}\normalsize
	\caption{Accuracy of SSH Framework and SRP (Sign Random Projection) on ECG and Random Walk Dataset for retrieving top-$k$ ($k=5, 10, 20, 50$) time series. We can see that SRP performs very poorly due to lack of alignment. The proposed SSH Framework on the other hand is significantly accurate. We can conclude that alignment is critical for these datasets.}
	\label{tab:acc}
	\begin{center}
		\renewcommand{\arraystretch}{1.0}
		\begin{tabular} {ccccccc}
			\hline\noalign{\smallskip}
			\textbf{Dataset} & \textbf{Time Series Length} & \textbf{Method} & \textbf{Top-5} & \textbf{Top-10} & \textbf{Top-20} & \textbf{Top-50} \\
			\noalign{\smallskip}
			\hline
			\noalign{\smallskip}
			\multirow{8}*{\centering{ECG}} & \multirow{2}*{\centering{128}} &
			SSH & $\mathbf{1.00 \pm 0.00}$ & $\mathbf{1.00 \pm 0.00}$ & $\mathbf{0.95 \pm 0.05}$ & $\mathbf{0.90 \pm 0.02}$ \\
			\cline{3-7}
			~&~& SRP & $0.20 \pm 0.00$ & $0.10 \pm 0.10$ & $0.10 \pm 0.05$ & $0.04 \pm 0.02$ \\
			\cline{2-7}
			~ & \multirow{2}*{\centering{512}}
			& SSH & $\mathbf{1.00 \pm 0.00}$ & $\mathbf{1.00 \pm 0.00}$ & $\mathbf{0.90 \pm 0.05}$ & $\mathbf{0.88 \pm 0.02}$ \\
			\cline{3-7}
			~&~& SRP & $0.00 \pm 0.00$ &  $0.10 \pm 0.10$ &  $0.05 \pm 0.05$ &  $0.04 \pm 0.02$ \\
			\cline{2-7}
			~ & \multirow{2}*{\centering{1024}}
			& SSH & $\mathbf{1.00 \pm 0.00}$ & $\mathbf{1.00 \pm 0.00}$ & $\mathbf{0.95 \pm 0.05}$ & $\mathbf{0.92 \pm 0.02}$ \\
			\cline{3-7}
			~&~& SRP &  $0.00 \pm 0.00$ & $0.00 \pm 0.00$ & $0.05 \pm 0.05$ &  $0.02 \pm 0.02$ \\
			\cline{2-7}
			~ & \multirow{2}*{\centering{2048}}
			& SSH & $\mathbf{1.00 \pm 0.00}$ & $\mathbf{1.00 \pm 0.00}$ & $\mathbf{0.95 \pm 0.05}$ & $\mathbf{0.94 \pm 0.02}$ \\
			\cline{3-7}
			~&~& SRP &  $0.00 \pm 0.00$ &  $0.00 \pm 0.00$ &  $0.00 \pm 0.00$ &  $0.00 \pm 0.00$ \\
			\hline
			\multirow{8}*{\centering{Random Walk}} & \multirow{2}*{\centering{128}} &
			SSH & $\mathbf{1.00 \pm 0.00}$ & $\mathbf{1.00 \pm 0.00}$ & $\mathbf{0.95 \pm 0.00}$ & $\mathbf{0.88 \pm 0.02}$ \\
			\cline{3-7}
			~&~& SRP &  $0.00 \pm 0.00$ &  $0.20 \pm 0.10$ &  $0.10 \pm 0.05$ &  $0.04 \pm 0.02$ \\
			\cline{2-7}
			~ & \multirow{2}*{\centering{512}}
			& SSH & $\mathbf{1.00 \pm 0.00}$ & $\mathbf{1.00 \pm 0.00}$ & $\mathbf{0.95 \pm 0.00}$ & $\mathbf{0.86 \pm 0.02}$ \\
			\cline{3-7}
			~&~& SRP &  $0.00 \pm 0.00$ &  $0.00 \pm 0.00$ & $0.05 \pm 0.00$ &  $0.04 \pm 0.02$ \\
			\cline{2-7}
			~ & \multirow{2}*{\centering{1024}}
			& SSH & $\mathbf{1.00 \pm 0.00}$ & $\mathbf{1.00 \pm 0.00}$ & $\mathbf{0.90 \pm 0.10}$ & $\mathbf{0.90 \pm 0.04}$\\
			\cline{3-7}
			~&~& SRP &  $0.00 \pm 0.00$ &  $0.10 \pm 0.00$ &  $0.05 \pm 0.05$ &  $0.04 \pm 0.00$ \\
			\cline{2-7}
			~ & \multirow{2}*{\centering{2048}}
			& SSH & $\mathbf{1.00 \pm 0.00}$ & $\mathbf{1.00 \pm 0.00}$ & $\mathbf{0.95 \pm 0.05}$ & $\mathbf{0.92 \pm 0.02}$ \\
			\cline{3-7}
			~&~& SRP & $0.00 \pm 0.00$ & $0.00 \pm 0.00$ & $0.00 \pm 0.00$ & $0.02 \pm 0.02$ \\
			\hline
		\end{tabular}
	\end{center}
\end{table*}

\begin{table*}[t]\normalsize
	\caption{CPU Execution time (in seconds) of UCR Suite and our proposed hashing method on ECG and Random Walk Dataset, with increasing query length. Hashing algorithm is consistently faster than UCR suite and gets even better with increase in query length. For longer query time series, hashing can be 20x faster.}
	\label{time}
	\begin{center}
		\renewcommand{\arraystretch}{1.2}
		\begin{tabular} {cccccc}
			\hline\noalign{\smallskip}
			\textbf{Dataset} & \textbf{Method} & \textbf{128} & \textbf{512} & \textbf{1024} & \textbf{2048} \\
			\noalign{\smallskip}
			\hline
			\noalign{\smallskip}
			\multirow{2}*{\centering{ECG}}
			& SSH & \textbf{2.30024}  & \textbf{5.50103} & \textbf{39.578} & \textbf{339.57}     \\
			\cline{2-6}
			& Branch-and-Bounds (UCR Suite)&  7.90036  & 20.2823 &  309.578 & 7934.615     \\
			\hline
			\multirow{2}*{\centering{Random Walk}}
			& SSH & \textbf{1.21002}  & \textbf{3.20156} & \textbf{15.2061} & \textbf{216.48035}   \\
			\cline{2-6}
			& Branch-and-Bounds (UCR Suite) &  3.32005  & 42.12  &  297.652 & 1934.615     \\
			\hline			
		\end{tabular}
	\end{center}
\end{table*}

\begin{table*}[t]
	\caption{Percentage of time series filtered by the SSH for different query length. Hashing, unlike branch and bound, becomes more effective for longer sequences.}
	\label{filtered}
	\begin{center}
		\renewcommand{\arraystretch}{1.2}
		\begin{tabular} {cccccc}
			\hline\noalign{\smallskip}
			\textbf{Dataset} & \textbf{Method} & \textbf{128} & \textbf{512} & \textbf{1024} & \textbf{2048} \\
			\noalign{\smallskip}
			\hline
			\noalign{\smallskip}
			\multirow{2}*{\centering{ECG}}
			& SSH Algorithm (Full) & \textbf{99.9\%}  & \textbf{98.8\%} & \textbf{90.8\%} & \textbf{95.7\%}     \\
			\cline{2-6}
			& Pruned by Hashing alone (SSH) & 72.4\%  & 76.4\% & 88.7\% & 95.4\%     \\
			\cline{2-6}
			& Branch-and-Bounds (UCR Suite)&  99.7\% & 94.96\% &  18.70\% & 7.76\%     \\
			\hline
			\multirow{2}*{\centering{Random Walk}}
			& SSH Algorithm (Full) & \textbf{99.6\%}  & \textbf{97.6\%} & \textbf{94.2\%} & \textbf{92.6\%}   \\
			\cline{2-6}
			& Pruned by Hashing alone(SSH) & 75.4\%  & 86.4\% & 91.7\% & 92.4\%    \\
			\cline{2-6}
			& Branch-and-Bounds (UCR Suite) &  98.6\%  & 82.7\%  &  30.2\% & 3.5\%     \\
			\hline			
		\end{tabular}
	\end{center}
\end{table*}

\subsection{Datasets}
To evaluate the effectiveness of our method for searching over time series, we choose two publicly available large time series data which were also used by the UCR suite paper \cite{rakthanmanon2012searching}: Random Walk, and ECG \footnote{http://www.cs.ucr.edu/~eamonn/UCRsuite.html}.
Random Walk is a benchmark dataset, which is often used for testing the similarity search methods~\cite{rakthanmanon2012searching}.
The ECG data consists of 22 hours and 23 minutes of ECG data (20,140,000 data points).

We also processed both the datasets in the same manner as suggested in~\cite{rakthanmanon2012searching}.
The process is as follow:
Given the very long time series $S=(s_1,s_2,...,s_m)$, and a time series length $t$ ($t=128,~512,~1024,~2048$).
We extract a time series data set $D=\{S^i| 1<i<m-t+1\}$, where each $S^i$ in $D$ is denoted as:
\[
S_i = (s_i,s_i+1,...,s_{i+t-1})
\]

After we get the time series data set $D=\{S^i| 1<i<m-t+1\}$, we can then do the similarity searching tasks.

For our proposed SSH method, the choices of $W$, $\delta$ and $n$ were determined using a holdout sample.

For the 22 Hour ECG time series data. We choose window size $W$ = 80, $\delta = 3$, and $n = 15$ for n-gram based set construction, and we use $20$ hash tables, for each hash table using our hash as index.
For the Random Walk Benchmark time series data. We choose window size $W = 30$, $\delta = 5$, and  $n = 15$ for n-gram based set construction, and we use $20$ hash tables.  The effect and choice of these values are explained in Section~\ref{sec:parastudy}.

%Random Walks is a benchmark dataset, which is often used for testing the similarity search methods~\cite{rakthanmanon2012searching}.
%In this paper, we use the random walks data set provided by UCR suite.
%This data set is ideal for evaluating sub-series search problem. It contains some query time series, and one very long time series ($20$ million data point).
%Following~\cite{rakthanmanon2012searching}, we transfer this data set into a time series similarity search data set. We generate all the sub-series with different length (e.g. 128, 512, 1024, 2048) in that data set. Our final data set consists $20$ million time series.
%
%The second data set we use is the 22 hours and 23 minutes of ECG data (20,140,000 data points) together with the exact query. This data set was also used as a  benchmark in the UCR suite paper. Again following~\cite{rakthanmanon2012searching}, we generate all the sub-series from this very long time series(20,140,000 datapoints time series) for evaluations.

\subsection{Baselines}

Since UCR suite is the state-of-the-art algorithm for searching over time series data, we use it as our best branch-and-bound baseline.
Note that branch-and-bound baselines are exact.

We point out here that there is no known data independent hashing scheme for time series data that can handle misalignment of time series. So, as another sanity check, we also compare the performance of vanilla hashing scheme the signed random projections ({\bf SRP}) to confirm if the alignment is a critical aspect.  For SRP, we simply regard the time series as long vectors. If alignment is not critical then treating time series as vectors is a good idea and we can expect SRP to perform well.

\subsection{Accuracy and Ranking Evaluation}

{\bf Task:} We consider the standard benchmark task of near-neighbor search over time series data with DTW as the gold standard measure.

To understand the variance of performance we run our hashing algorithm SSH, as described in Section~\ref{algorithm}, for searching top-$k$ near neighbors. The gold standard top-$k$
neighbors were based on the actual DTW similarity. For the SRP baseline, we replace SSH indexing procedure with SRP hash function. For a rigorous evaluation we run these algorithms with a different values of  $k = \{5, \ 10, \ 20, \ 50\}$.

{\bf Evaluation Metric}
We use precision and NDCG (Normalized Discounted Cumulative Gain) \cite{wang2013theoretical} to evaluate the accuracy and rankings of our method.

Precision for a search result is defined as
\begin{equation*}
	Precision = \frac{relevant seen}{k},
\end{equation*}
here relevant seen denotes the number of top-$k$ gold standard time series returned by the algorithm.

Just observing the precision is not always a good indicator of rankings. We also evaluate the rankings of the top-$k$ candidates reported by our algorithm with the rankings generated by DTW measure. We use NDCG (Normalized Discounted Cumulative Gain), which is a widely used ranking evaluation metric. NDCG  \cite{wang2013theoretical} is defined as:
\begin{equation*}
	nDCG = \frac{DCG}{IDCG},
\end{equation*}
where DCG can be calculated as $DCG = \sum_{i=1}^{k} \frac{R_i}{\log_2(i)}$, and IDCG denotes the DCG for the ground truth ranking result. $R_i$ denotes the graded relevance of the result at position $i$. In this experiment, $R_i$ is calculated as $R_i = k - i.$

{\bf Result:} The ranking result on ECG and Random walk dataset is shown in Figure. \ref{fig:ndcg}. We summarize the precision of different methodologies in Table. \ref{tab:acc}. Note, UCR suite is an exact method so it will always have an accuracy of 100\% and NDCG is always 1.

We can see from Figure. \ref{fig:ndcg} that the proposed SSH based ranking achieves near perfect NDCG value for most values of $k$ and gracefully decreases for large $k$. This deterioration with increasing $k$ is expected as hashing techniques are meant for high similarity region. On the contrary, the performance of SRP is quite poor irrespective of the values of $k$, indicating the importance of alignment.

In Table. \ref{tab:acc}, we can see that for the top-$5$ and top-$10$ similarity search tasks, our proposed method can get 100\% accuracy for both the benchmark datasets. For large $k \ge 20$ we see some loss in the accuracy.  As expected, hashing based methods are very accurate at high similarity levels which are generally of interest for near-neighbor search.  The performance of SRP, as expected,  is very poor indicating the need for alignment in time series data. The success of our method clearly shows that our proposal can get the right alignment.   We can clearly see that despite SSH being the approximate method the impact of approximation is negligible on the final accuracy.
The accuracy trends are consistent for both the datasets.

%We now show the speedup obtained due to the SSH procedure.

\subsection{Speed Comparison}
We now demonstrate the speedup obtained using the SSH procedure.  We compute the average query time which is the time required to retrieve top-k candidates using Algorithm~\ref{algorithm}.  The query time includes the time needed to compute the SSH indexes of the query.

The CPU execution time of our method and exact search method using UCR Suite is shown in Table \ref{time}. We can clearly see that hashing based method is significantly faster in all the experiments consistently over both the data sets irrespective of the length of the query.

It can be clearly seen from the table that when the query length increases, the UCR suite scales very poorly. For searching with long time series (e.g. 2048 or higher) hashing based method is drastically efficient compared to the UCR Suite. It can be around 20 times faster than the UCR suite.

To understand the effectiveness of hashing in pruning the search space, we also show the number of time series that were filtered by our proposed algorithm. We also highlight the candidates pruned by hashing alone and separate it from the total candidates pruned which include additional pruning in step 10 of Algorithm~\ref{HaTSQue}. We summarize these percentages in Table. \ref{filtered}. Hashing itself prunes down the candidates drastically. For shorter queries, it is advantageous to use branch-and-bound to filter further the candidates returned by hashing. As expected for longer queries hashing is sufficient, and the additional branch and bound pruning by our algorithm leads to negligible advantages. Hashing based pruning works even better with an increase in the time series length.  This is expected because hashing based method is independent of dimensions and only pick time series with high similarity.   These numbers also demonstrate the power of our method when dealing with long time series.

It should be further noted that hashing based filtering, unlike branch and bound, does not even require to evaluate any cheap lower bound and thus are truly sub-linear. In particular, branch and bound prune a candidate by computing a cheap lower bound which still requires enumerating all the time series. Hashing eliminates by bucketing without even touching the pruned candidates.

\subsection{Parameter Study}
\label{sec:parastudy}

As we introduced in Section \ref{sec:practical}.
The proposed hashing scheme takes $W$ (the dimension of the filter $r$), $\delta$ (the shift size) and $n$ (shingle length) as parameters. The choice of these parameters is critical for the performance. In this section, we shall show the impact of these three parameters on the retrieving accuracy and execution time. This study will also explain the selection procedure of the parameters we used in our experiments.

\subsubsection{$W$ (the dimension of filter)}

The dimension $W$ of the filter $r$ in our framework is a critical parameter.
%The choice of $W$ impacts the accuracy and execution time of our framework.
If $W$ is too large, then the 1-bit sketch is likely to be non-informative and won't capture temporal trends.  Also, it may merge significant patterns of the time series. On the other hand, if the choice of $W$ is too small, then the sub-series may only contain component information which can be very noisy.  Thus there is a trade-off.
From the execution time perspective view, if we choose large $W$, the preprocessing execution time will increase due to the inner product operation.  As a result, proper choice of $W$ is an imperative step of our framework.

\begin{figure}[t]
	\centering
	\includegraphics[width=0.45\textwidth]{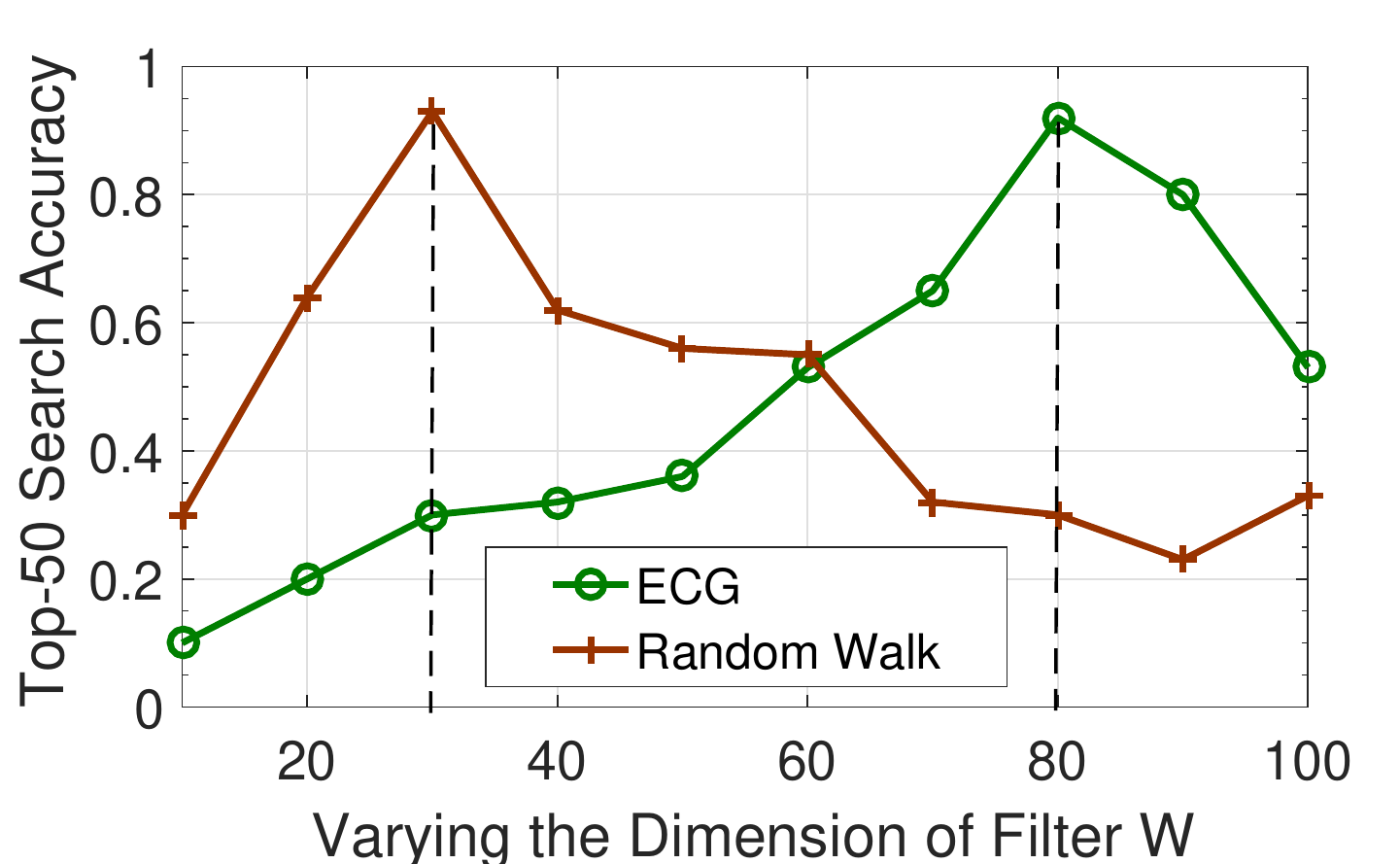}
	\caption{Accuracy with respect to the filter dimension $W$ for the two data sets.}
	\label{para:w:a}
\end{figure}

\begin{figure}[t]
	\centering
	\includegraphics[width=0.45\textwidth]{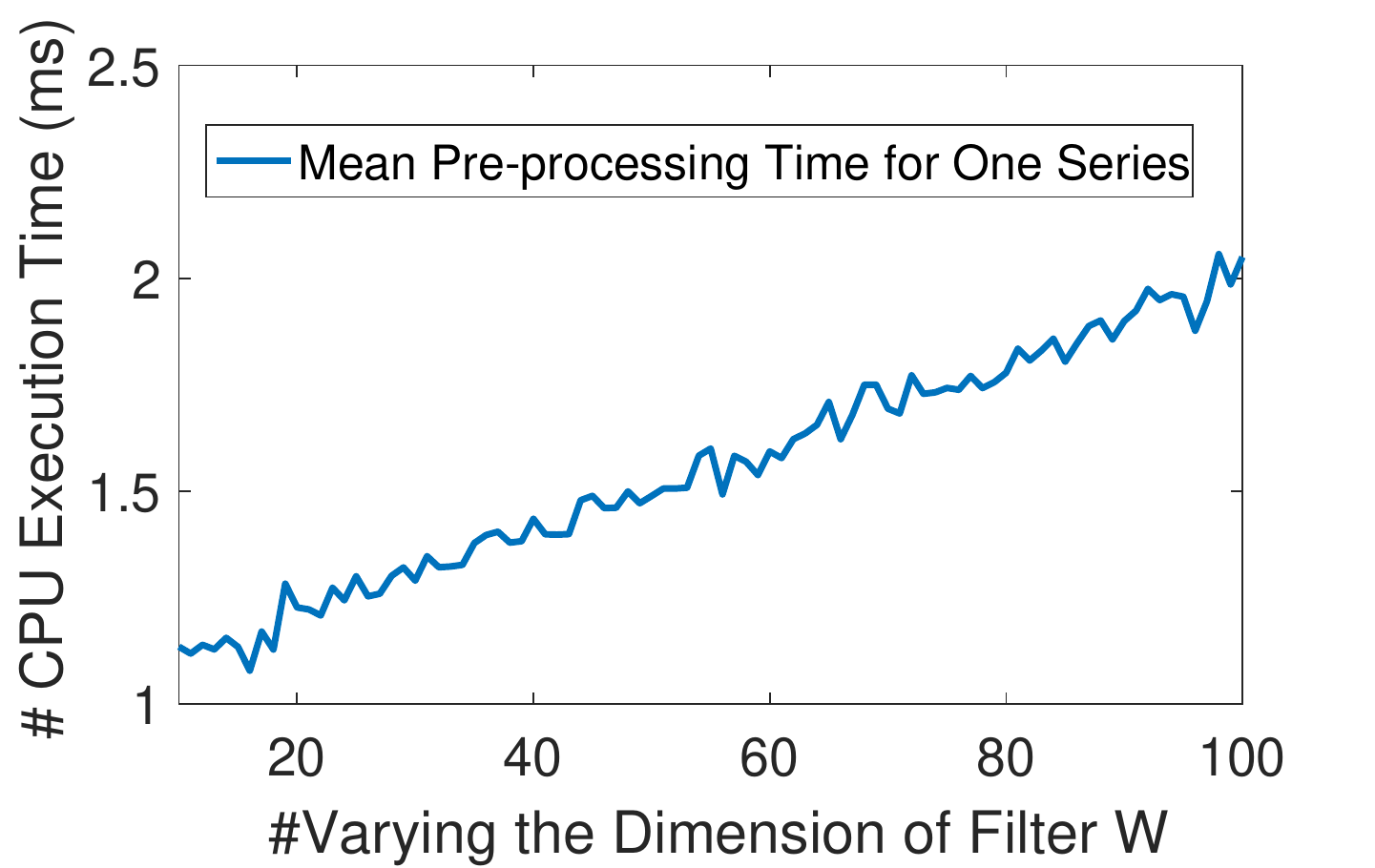}
	\caption{Preprocessing time with respect to the filter dimension $W$.}
	\label{para:w:e}
\end{figure}

Fig. \ref{para:w:a} shows the precision of SSH indexes, for top-50 neighbors, with varying filter dimension $W$ for the two data sets.
We can see from the figure that when the $W$ is small, the accuracy is reduced. With the increase of the filter dimension $W$, the accuracy starts increasing and after reaching a sweet spot drops again.
We achieve the sweet spot at $W$=80 for ECG time series data and $W$=30 for random walk data respectively.

Fig. \ref{para:w:e} shows the preprocessing time with varying filter dimension $W$ for the two data sets.
From the result, we can see that the average running time for preprocessing on a single time series is linear to the dimension of $W$.
This is because the preprocessing time will increase due to the increase in the number of inner product operation.

\subsubsection{$\delta$ (the shift size)}

\begin{figure}[t]
	\centering
	\includegraphics[width=0.45\textwidth]{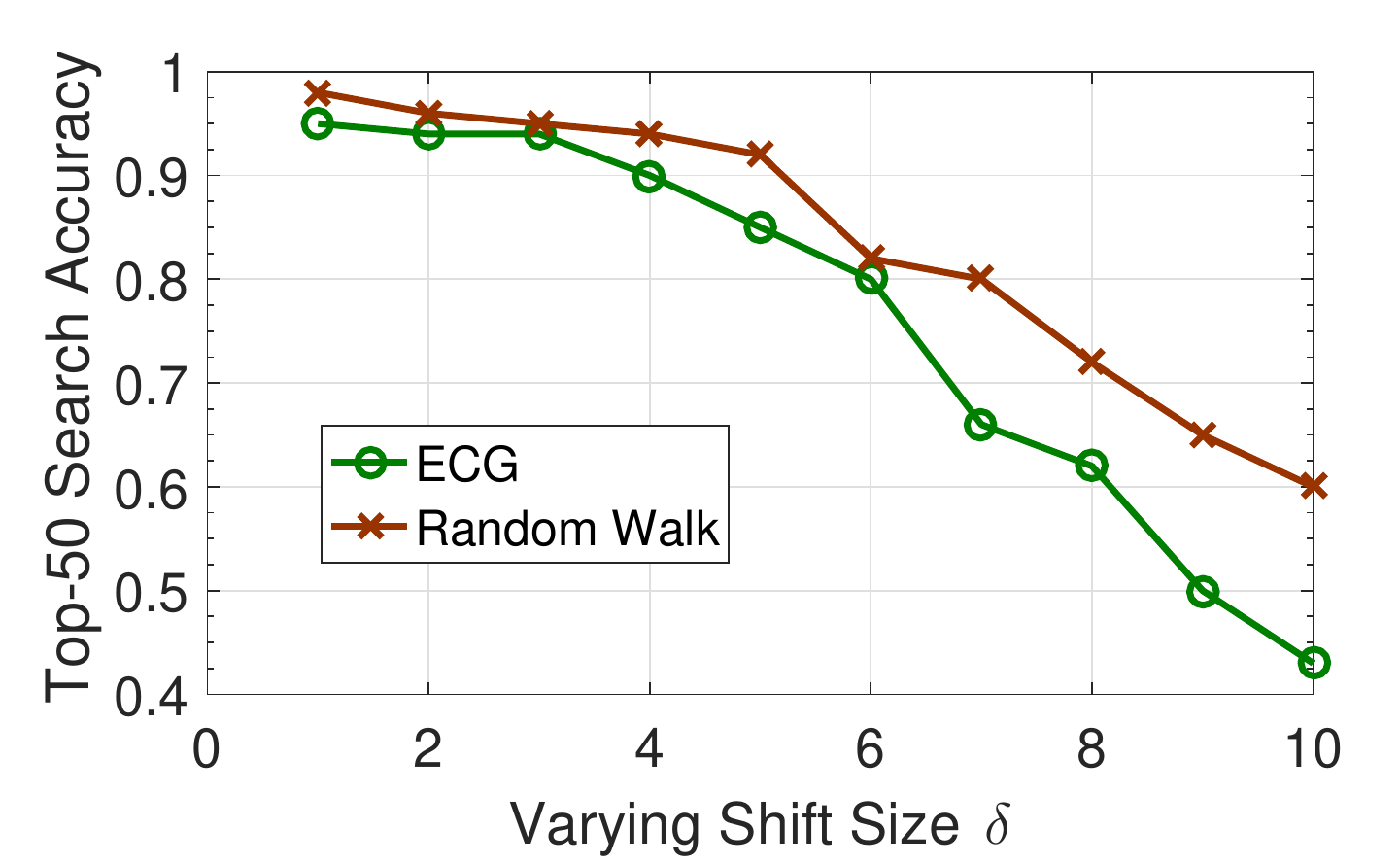}
	\caption{Accuracy with respect to the shift size $\delta$.}
	\label{para:d:a}
\end{figure}

\begin{figure}[t]
	\centering
	\includegraphics[width=0.45\textwidth]{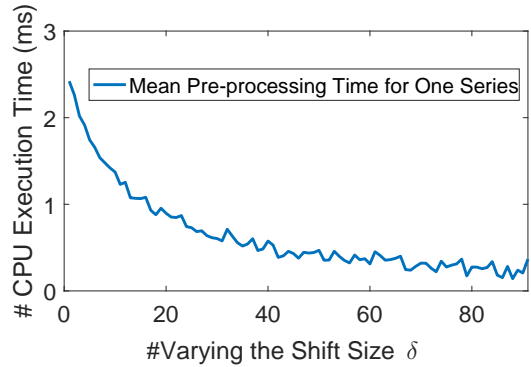}
	\caption{Preprocessing time with respect to the shift size $\delta$}
	\label{para:d:e}
\end{figure}

The shift size $\delta$ of the SSH scheme shows a consistent trend of decreasing the accuracy with an increase in $\delta$.
The best $\delta$ is $\delta=1$.  However, a smaller $\delta$  increase the execution complexity of SSH because of the number of inner products.
Large $\delta$ leads to information loss.
Fig. \ref{para:d:a} shows the accuracy with the shift size $\delta$ for the two data sets, whereas Fig. \ref{para:d:e} shows the preprocessing time by varying the shift size $\delta$ for the two data sets.   From the result, we can see that the average running time for preprocessing increases with the decreasing of the shift size.
To balance this accuracy-time trade-off we chose $\delta=3$ for ECG and $\delta=5$ for random walk data respectively.

\subsubsection{$n$ (shingle length)}

\begin{figure}[t]
	\centering
	\includegraphics[width=0.45\textwidth]{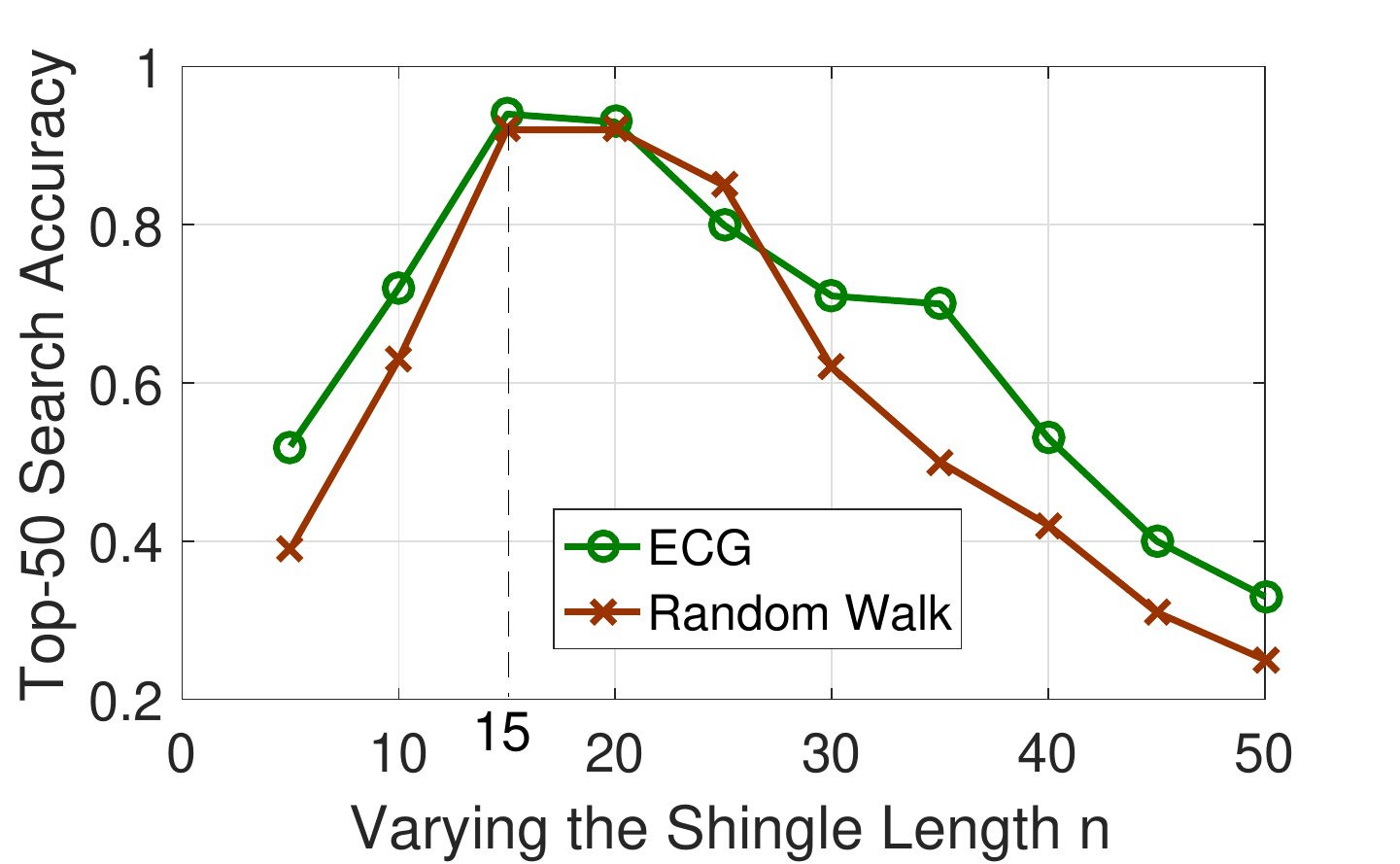}
	\caption{Accuracy by varying the shingle length $n$ for the two data sets.}
	\label{para:n:a}
\end{figure}

\begin{figure}[t]
	\centering
	\includegraphics[width=0.45\textwidth]{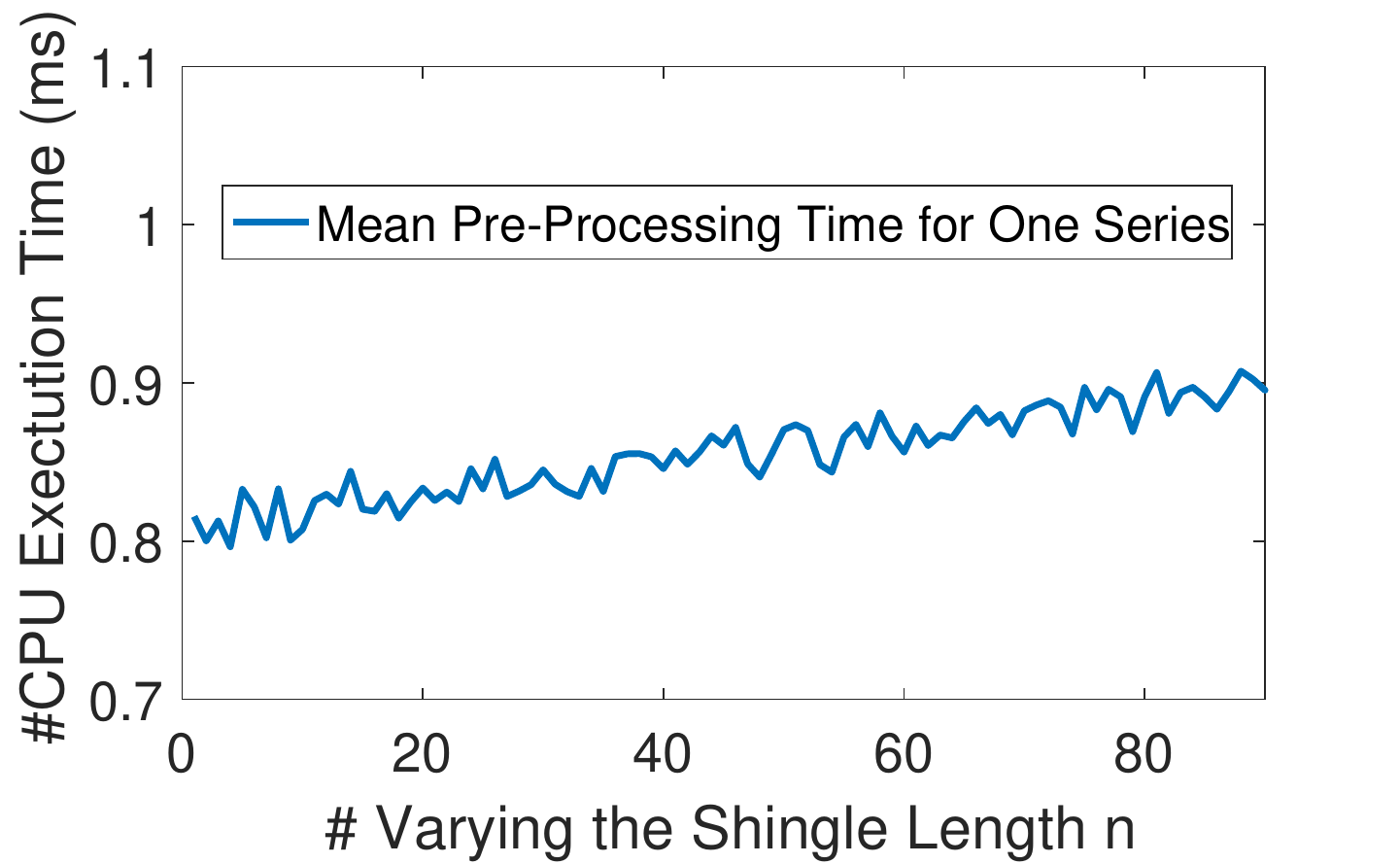}
	\caption{Preprocessing execution time by varying the filter dimension $n$ for the two data sets.}
	\label{para:n:e}
\end{figure}

The shingle length $n$ in SSH turns out to be a sensitive and critical parameter. Just like the behavior of $n$-grams in the text,  too large to too little $n$ hurts the performance.

Fig. \ref{para:n:a} shows the accuracy by varying the Shingle length $n$ for the two data sets.  We can see from the figure that when the $n$ is too small, the accuracy is poor.  With the increasing of the shingle length $n$, the accuracy also increase.
For both ECG and Random Walk datasets, $n$=15 seems to be the right sweet spot.  With further increasing in $n$, the accuracy start to decrease.

Fig. \ref{para:n:e} shows the preprocessing execution time by varying the filter dimension $n$ for the two data sets.
As expected, we can see that the average running time for preprocessing is linear to the dimension of $n$.
When the shingle length $n$ increases, the constructed weighted set $S$ will become larger, thus the execution time will also increase.

\section {Conclusions}

%{
%%\color{red}
%Similarity search with DTW is a popular but expensive data mining task.
%Speeding up similarity search with DTW is an important research direction.
%Branch and bound based candidate pruning was the most popular method for improving search efficiency with DTW.
%However, branch-and-bound techniques suffer from the curse of dimensionality.
%
%Therefore, in this paper, we proposed SSH (Sketch, Shingle \& Hash), the first data independent hashing scheme which does both the alignment and matching on time series data. Unlike branch and bound based approaches our scheme does not deteriorate with an increase in the length of query time series.
%SSH combines carefully chosen three step procedure for indexing time series data which as we demonstrate are ideal for searching with DTW similarity measure.
%For similarity search with time series data, we show around 20x speedup over the fastest package UCR suite on two benchmark datasets.
%
%}

DTW is a widely popular but expensive similarity measure.
Speeding up DTW is an important research direction.
Branch and bound based candidate pruning was the most popular method for improving search efficiency with DTW.   However, branch-and-bound techniques suffer from the curse of dimensionality.
Our proposed framework provides an alternative route of randomized indexing to prune the candidates more efficiently.

We have proposed SSH (Sketch, Shingle \& Hash) the first indexing scheme which does both the alignment and matching on time series data.  Unlike branch and bound based approaches our scheme does not deteriorate with an increase in the length of query time series are is free from the curse of dimensionality.

SSH combines carefully chosen three step procedure for indexing time series data which as we demonstrate are ideal for searching with DTW similarity measure.   For similarity search with time series data, we show around 20x speedup over the fastest package UCR suite on two benchmark datasets.

\section*{Acknowledgments}

We are thankful for UCR time series Team for providing the UCR Suite code and the corresponding time series data. This work was supported by Rice Faculty Initiative Award 2016.

\balance

\bibliographystyle{abbrv}

\bibliography{www}

%\balancecolumns % GM June 2007
% That's all folks!
\end{document}